\documentclass[final,5p,times,twocolumn]{elsarticle}

\usepackage{color}
\usepackage[fleqn]{amsmath}
\usepackage{amssymb}
\usepackage{subfigure}
\usepackage{graphicx}
\usepackage{caption}
\usepackage{picins}
\usepackage{wrapfig}
\usepackage{booktabs}
\usepackage{multirow}
\biboptions{numbers,sort&compress}
\begin{document}
\captionsetup[figure]{labelfont={bf},labelformat={default},labelsep=period,name={Fig.}}
\begin{frontmatter}

\title{A Novel Dual Dense Connection Network for Video Super-resolution}
%\tnotetext[label1]{This paper is supported by the National Natural Science Foundation of China(No. 11571325) and the Fundamental Research Funds for the Central Universities(No. CUC2019 A002).}

\author[1]{Guofang Li}
\affiliation[1]{organization={School of Information and Communication Engineering,  Communication University of China},
             addressline={},
             city={Beijing},
             postcode={100024},
             country={China}}

\author[2]{Yonggui Zhu\corref{cor1}}
\cortext[cor1]{Corresponding author}
\affiliation[2]{organization={School of Data Science and Intelligent Media, Communication University of China},
             addressline={},
             city={Beijing},
             postcode={100024},
             country={China}}
\ead{ygzhu@cuc.edu.cn}
\begin{abstract}
%% Text of abstract
Video super-resolution (VSR) refers to the reconstruction of high-resolution (HR) video from the corresponding low-resolution (LR) video. Recently, VSR has received increasing attention. In this paper, we propose a novel dual dense connection network that can generate high-quality super-resolution (SR) results. The input frames are creatively divided into reference frame, pre-temporal group and post-temporal group, representing information in different time periods. This grouping method provides accurate information of different time periods without causing time information disorder. Meanwhile, we produce a new loss function, which is beneficial to enhance the convergence ability of the model. Experiments show that our model is superior to other advanced models in Vid4 datasets and SPMCS-11 datasets.
\end{abstract}

%%Graphical abstract
%%\begin{graphicalabstract}
%\includegraphics{grabs}
%%\end{graphicalabstract}

%%Research highlights
%\begin{highlights}
%\item Research highlight 1
%\item Research highlight 2
%\end{highlights}


\end{frontmatter}

%% \linenumbers

%% main text
\section{Introduction}
Super-resolution refers to yielding HR images from the corresponding LR images. With the increasing quality of videos, high-definition, ultra high-definition and even Blue-ray videos have become a part of people's life. In this case, SR technology which can amplify resolution has received further attention and development. Up to now, SR has been extensively used in face recognition \cite{face-recognition}, video surveillance \cite{video-surveillance}, medical imaging \cite{medical-imaging} and other fields.

In order to obtain high quality images, studies have proposed numerous effective methods. Initially, researchers utilize interpolation methods to obtain HR videos \cite{interpolation1, interpolation2}. These methods possess higher computing speed, but the results are poor. With the development of deep learning, the construction of SR model based on deep learning has become the mainstream research approach. Some meaning models have been built. For instance, Caballero et al. \cite{VSR-VESPCN} proposed spatio-temporal networks and motion compensation to construction SR models. Wang et al. \cite{VSR-EDVR} presented Pyramid, Cascading and Deformable (PCD) alignment module. Moreover, some researchers utilize recurrent neural networks to acquire time information between frames \cite{LSTM1, VSR-MMCNN}. A number of studies apply deformable convolution to further expand the field of feature extraction \cite{deformable2D, deformable3D}. Many studies use 3D convolution to directly fuse information from adjacent frames \cite{3D1, 3D2} and so on \cite{VSR-buchong1,VSR-PFNL,VSR-BasicVSR,VSR-buchong2}. However, these models can not perfectly solve the problems existing in SR, such as the existence of artifacts, large amount of model training, weak model generalization ability, etc. Thus, it is a very meaningful study to explore SR technology.

In all SR networks, densely connected convolutional network (DCCN) \cite{Densely_connected} is a famous idea in which each layer of the network accepts the output from all previous layers. However, with the network depth increases, the number of filters gradually expands, occupying a large amount of memory. In order to reduce the quantity of filters, the article \cite{Densely_connected} utilized $1\times1$ convolution to compress filters. Although this method reduced the number of filters, the information before compress could not be reused, resulting in information waste. Moreover, the paper \cite{Densely_connected1} repeatedly used DCCN. However, it could not solve the problem of rapidly increasing feature maps. Thus, in this paper, we propose a dual dense connection network (DDCN), which includes dual dense connection network based on 3D convolution (DDCN3D) and dual dense connection network based on 2D convolution (DDCN2D). The main purpose of DDCN is to build the inner and outer layer networks to alleviate the problem of excessive feature maps. In specific operations, the inner layer is responsible for extracting and compressing features, and the outer layer reconnects feature information to improve feature utilization.

\begin{figure*}[ht]
\begin{center}
\includegraphics[scale=0.6]{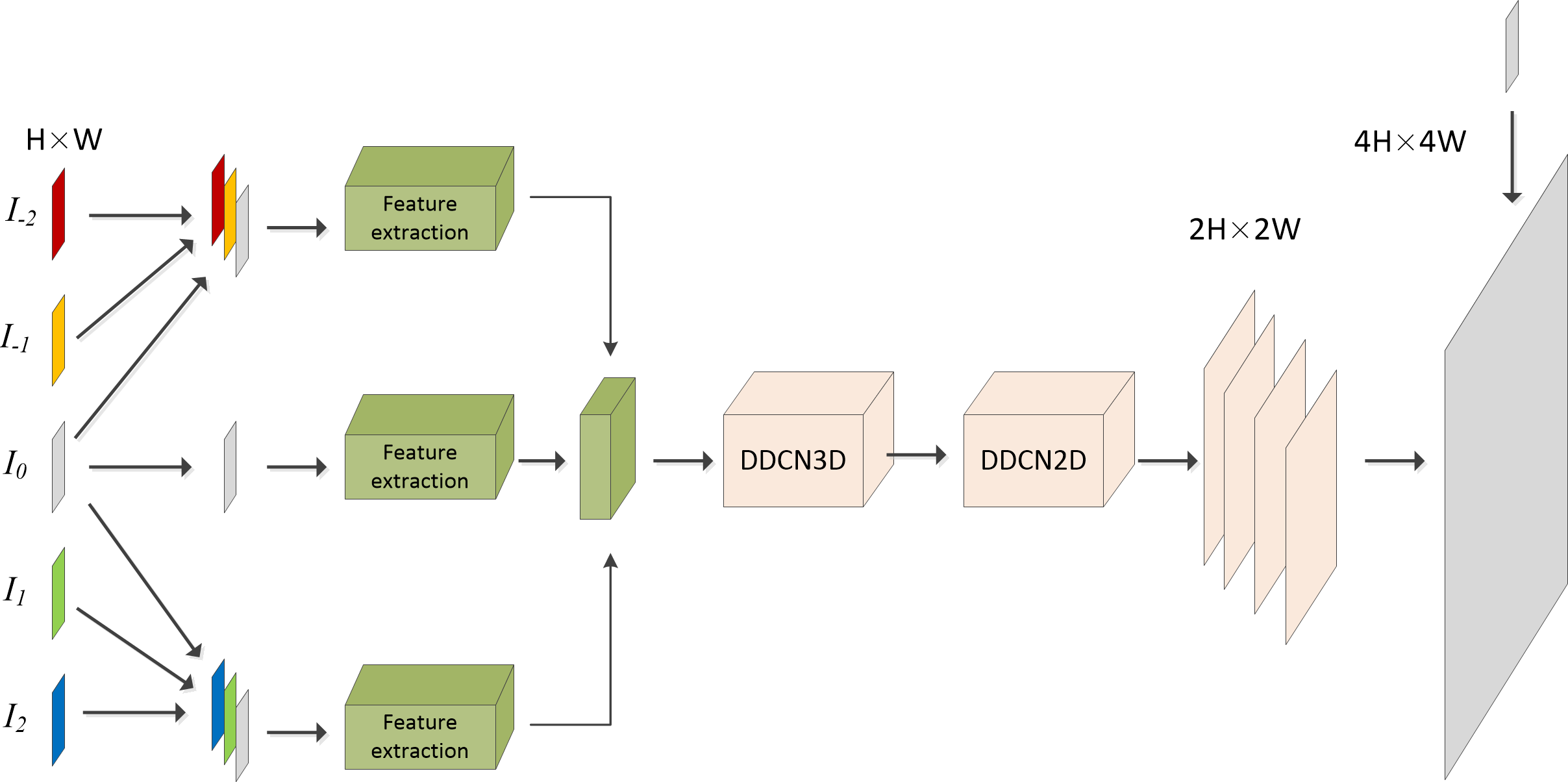}
\caption{Our model construction for 4$\times$ magnifiaction. $I_{-2}, \cdots, I_{2}$ indicates the input frame, where $I_0$ is reference frame. DDCN3D and DDCN2D refer to the DDCN network applying 3D convolution and the DDCN network applying 2D convolution respectively. H and W represent the height and width of the feature maps.}
\label{main_figure}
\end{center}
\end{figure*}

Moreover, we divide the input frames into pre-temporal group, reference frame and post-temporal group. Different groups obtain motion information in an implicit manner. This grouping method can effectively extract the information of the distant frame and ensure that offsets and artifacts are not generated. Meanwhile, since frames at different distances possess different values, we introduce the temporal attention to further enhance the ability of information extraction. Finally, we propose a novel loss function to replace the traditional regularization term and obtain high quality SR results.

The main contributions of this paper are as follows:

$\bullet$ We introduce a DDCN for video super-resolution, including inner layer and outer layer. DDCN can reduce the number of feature maps while ensuring the repeated use of features. In the specific operation, we utilize DDCN3D and DDCN2D to extract grouping features and reconstruction features.

$\bullet$ We come up with a new grouping method which consists of pre-temporal group, reference frame and post-temporal group. This grouping ensures the accuracy of information extraction.

$\bullet$ We present a new loss function, which can enhance the convergence ability of DDCN. The proposed method achieves advanced results on Vid4 datasets and SPMCS-11 datasets.
\section{Related work}
\subsection{Single Image Super Resolution}
With the widespread application of deep learning, SR has ushered in a new revolution. The SRCNN model proposed by Dong et al. \cite{ImageSR-SRCNN} was the first to apply deep learning to single image super resolution (SISR). They presented a three layers convolution neural network and achieved better effect. It was proved that deep learning possessed great potential in the field of SR. After this article, Kim et al. \cite{ImageSR-VDSR} came up with a very deep neural network and applied the residual network to the SR model, achieving better effect than SRCNN. Shi et al. \cite{ImageSR-SPM} presented sub-pixel magnification method, which mapped LR images to HR images without utilizing additional parameters. Liu et al. \cite{ImageSR-RFANet} proposed residual aggregation framework, which further applied the information of residual networks. Song et al. \cite{ImageSR-ADDerSR} came up with the idea of making use of additive neural network for SISR, which replaced the traditional convolution kernel multiplication operation in the calculation of output layer, saving numerous computation power.
\subsection{Video Super resolution}
VSR is an extension of SISR. In VSR, the temporal information between adjacent frames play a vital role. In order to acquire perfect results, studies have built a variety of modules. For instance, VESPCN \cite{VSR-VESPCN} applied the optical flow field which included coarse flow and fine flow to align adjacent frames, and constructed an end-to-end spatio-temporal module. Based on VESPCN, MMCNN \cite{VSR-MMCNN} combined optical flow field and long short-term memory to make more efficient use of inter-frame information and obtain more real details. TDAN \cite{deformable2D} was the first model to substitute the deformable convolution into VSR, which amplified the feature extraction ability of the model. EDVR \cite{VSR-EDVR} presented PCD alignment module, which changed the direct connection of LR images in TDAN. TMNet \cite{VSR-TMNet} designed temporal modulation block to modulate the PCD module. Meanwhile, TMNet conducted short-term and long-term feature fusion to better extract motion clues. The above models mainly apply 2D convolution. Experiments indicate that 3D convolution also possess significant values in VSR. For example, 3DSRNet \cite{3D2} utilized 3D convolution to construct the VSR model without motion alignment, maintaining the time depth of spatio-temporal feature maps, and extracted the temporal information between LR frames as much as possible. D3D \cite{deformable3D} proposed deformable 3D convolution with efficient spatio-temporal exploration and adaptive motion compensation capabilities. TGA \cite{3D1} presented the structure of intra-group fusion and inter-group fusion. And by complementing the characteristics of different group, TGA model yielded advanced results.

Although the above methods achieve excellent results, there are still some shortcomings. For instance, the model based on the optical-flow method is larger, and it is prone to produce artifacts, which affect the visual effects. Models applying deformable convolutions can reconstruct high-quality HR images. However, these models require a long training time and consume a mass of resources. Thus, our paper utilizes an implicit alignment method to avoid artifacts that are easy to generate when displaying alignment. Meanwhile, we apply 3D convolution instead of deformable convolution to reduce resource consumption.
\section{Our Method}
\subsection{Overview}
Given a consecutive odd frames $I^L_{-T}, \cdots, I^L_0, \cdots, I^L_T$ as input, where $I^L_0$ is the reference frame. The goal is to obtain the HR images of $I^H_0$. Our proposed model is shown in \textbf{Fig. \ref{main_figure}}. This model supports the input of any number of odd frames. In the specific operation, we divide the input frames into three groups. After that, we extract features from three groups to obtain feature information of different periods. Then, we apply feature fusion module to fully gain spatio-temporal information of the input frames, and utilize feature reconstruction module to further enhance and magnify the fused features. Finally, the HR image $I^H_0$ is acquired by adding the generated map by the model and the bicubic upsampling of the reference frame $I^L_0$.
\begin{figure}[ht]
\begin{center}
\includegraphics[scale=0.4]{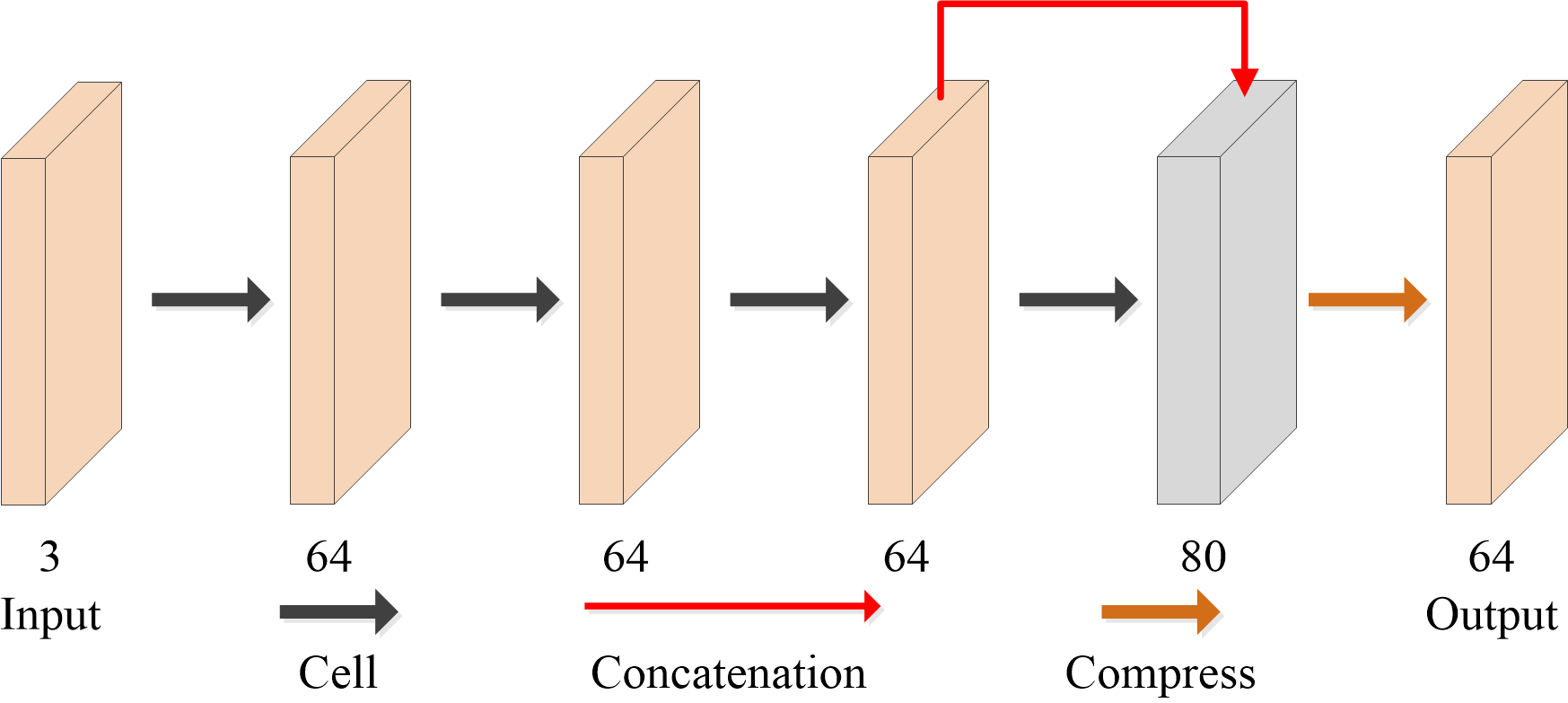}
\caption{Feature extraction of pre-temporal group and post-temporal group. The grey block indicates that the attention mechanism is applied at the current layer . The number is the quantity of feature maps for this layer.}
\label{feature_extraction_model}
\end{center}
\end{figure}
\subsection{Feature Extraction Module}
Feature extraction module consists of temporal grouping and feature extraction. For a consecutive odd frames $I^L_{-T}, \cdots, I^L_0, \cdots, I^L_T$, we divide frames into three groups, which include reference frame $I^L_0$, pre-temporal group $I^L_{-T}, I^L_{-T+1}, \cdots, I^L_0$ and post-temporal group $I^L_0, \cdots, I^L_{T-1}, I^L_T$. In the pre-temporal group, frames farther from the reference frame can capture more information, and frame closer from the reference frame can compensate for some missing details. The post-temporal group can be similarly explained. Moreover, the three groups represent the spatio-temporal information of past, present and future respectively. When we extract the features of the entire series of frames, this grouping method maintains the temporal consistency and avoids temporal confusion. At the same time, the reference frame $I^L_0$ is included in both pre-temporal group and post-temporal group to ensure that the extracted features will not emerge a large offset. For the reference frame, we utilize five $3\times3$ convolution to acquire feature. For the pre-temporal group and post-temporal group, we first construct a cell composed of a $1\times3\times3$ convolutional kernel, a batch normalization and a ReLU activation function. Then, we utilize four cells and temporal attention to gather information. Finally, a $1\times3\times3$ convolution kernel is applied to compress feature maps to 64. Our proposed model is displayed in \textbf{Fig. \ref{feature_extraction_model}}.
\subsection{Feature Fusion Module}
\begin{figure}[ht]
\begin{center}
\includegraphics[scale=0.35]{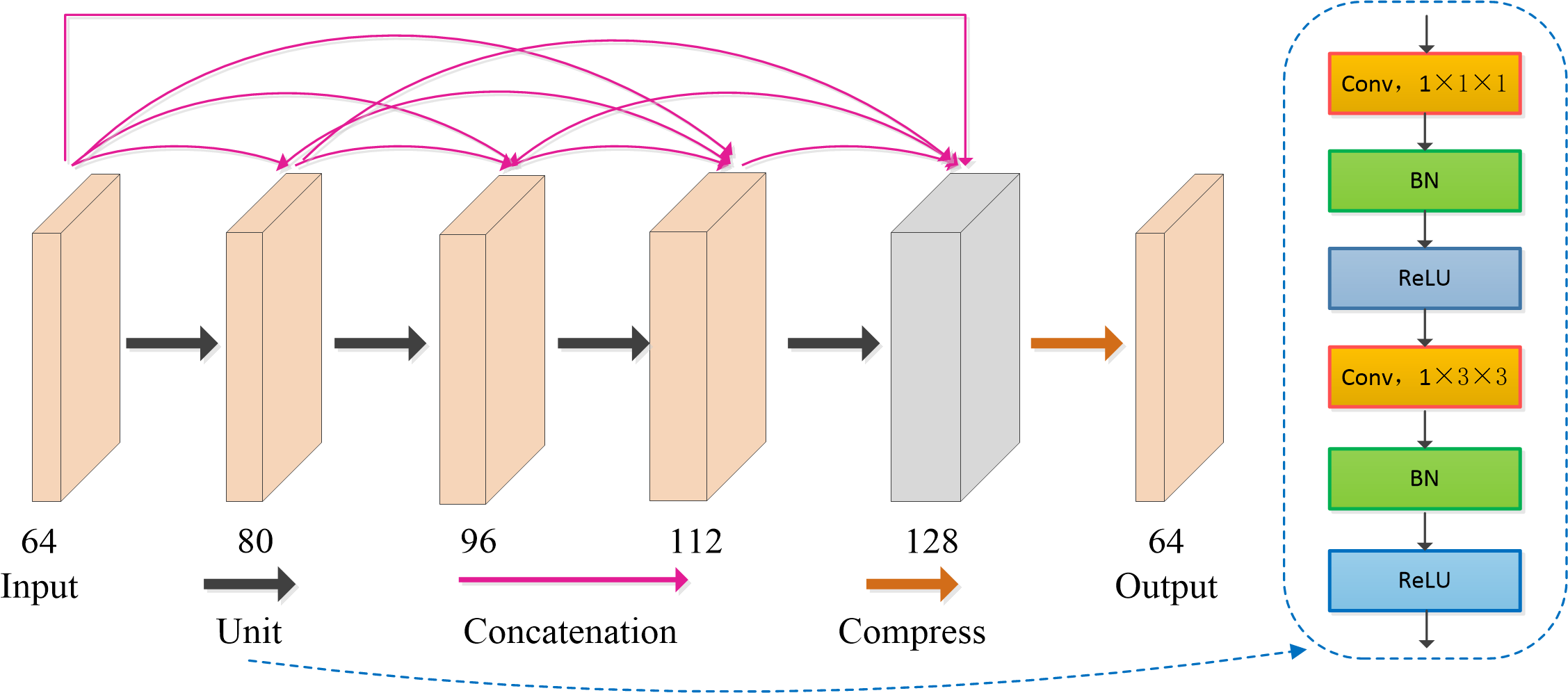}
\caption{The inner layer of DDCN3D. The grey block represents that the attention mechanism is applied at the current layer. The number indicates the quantity of feature maps for this layer.}
\label{inner_layer}
\end{center}
\end{figure}

\begin{figure}[ht]
\begin{center}
\includegraphics[scale=0.5]{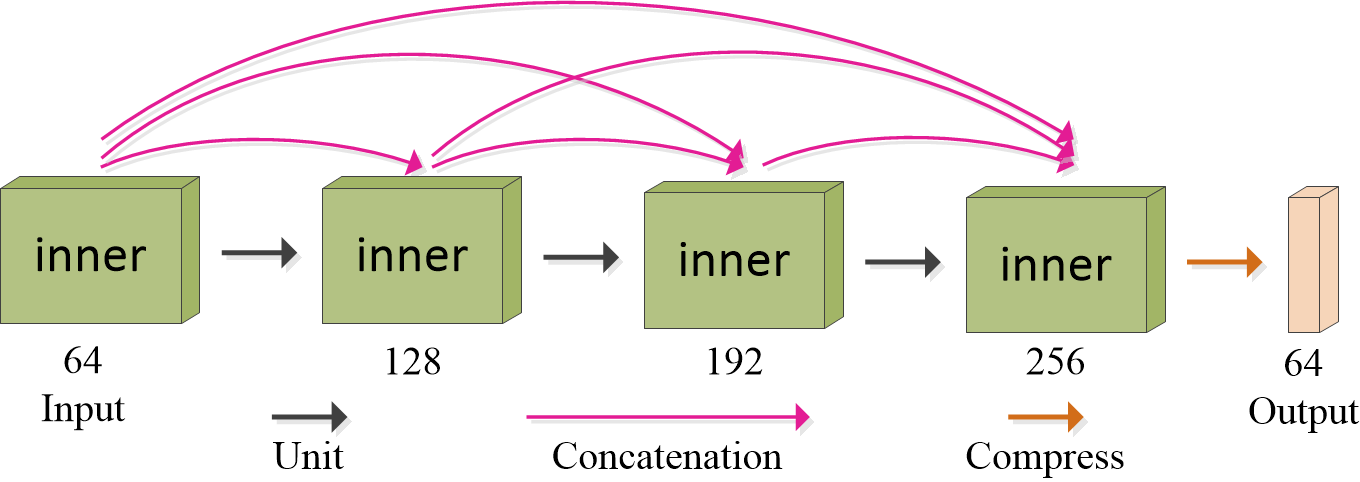}
\caption{The outer layer of DDCN3D and DDCN2D. The green blocks indicate the inner output from DDCN3D or DDCN2D. The number indicates the quantity of feature maps for this layer.}
\label{outer_layer}
\end{center}
\end{figure}

Feature fusion module consists of DDCN3D and fusion module. DDCN3D combines the idea of densely connected convolutional networks \cite{Densely_connected} and temporal attention mechanisms, and divides into inner-layer and outer-layer. The inspiration of the inner-layer design come from the inter-group fusion module in \cite{3D1}. The inner layer we designed is shown in \textbf{Fig. \ref{inner_layer}}, and the growth speed is set at 16. The outer-layer circulates the inner-layer in the form of dense connectivity, and the growth speed is set at 64. The flow chart is displayed in \textbf{Fig. \ref{outer_layer}}. Then, we fuse the output features of DDCN3D to obtain the spatio-temporal information of grouping features.

\subsection{Feature Reconstruction Module}
\begin{figure}[ht]
\begin{center}
\includegraphics[scale=0.4]{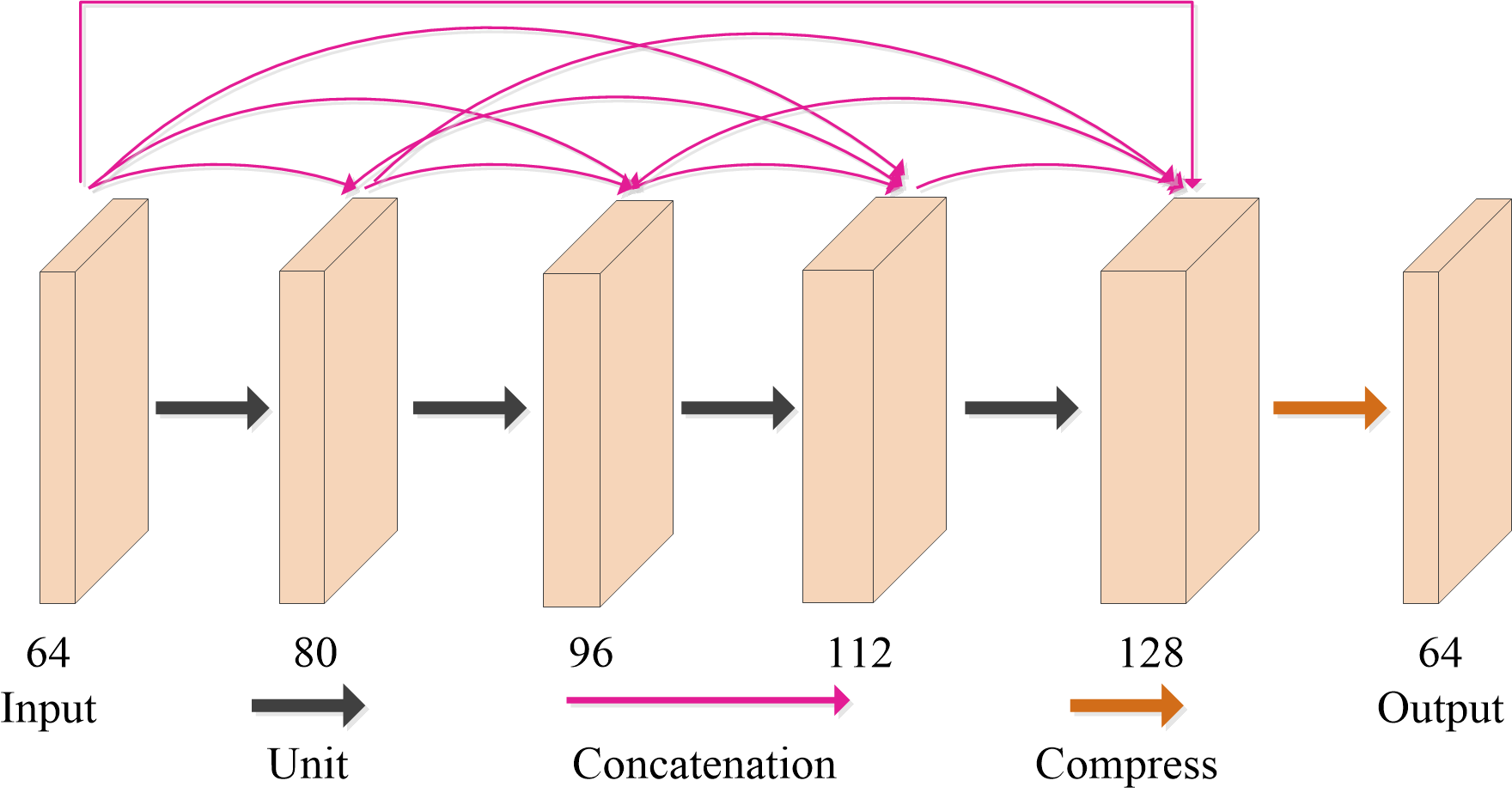}
\caption{The inner layer of DDCN2D. The number indicates the quantity of feature maps for this layer.}
\label{feature_reconstruction_model}
\end{center}
\end{figure}

Feature reconstruction module consists of DDCN2D and sub-pixel magnification. Similar to DDCN3D, DDCN2D is divided into inter-layer and outer-layer. The inner-layer is composed of a dense connection block and a compression block. The dense connection block consists of 4 units, and each unit contains a $1\times1$ convolution kernel and a $3\times3$ convolution kernel. Meanwhile, the compression block is a $3\times3$ convolution kernel. The flow chart is displayed in \textbf{Fig. \ref{feature_reconstruction_model}}. The outer-layer also circulates the inner-layer in the form of dense connection, as shown in \textbf{Fig. \ref{outer_layer}}. Then, we utilize sub-pixel magnification \cite{ImageSR-SPM} to obtain HR feature maps.
\subsection{Temporal Attention}
In order to better achieve feature extraction and feature fusion, we insert temporal attention into the two modules. In pre-temporal group and post-temporal group, the attention mechanism can analyze the importance of information in different frames. In the feature fusion module, different groups of information possess different values, and the attention mechanism can effectively extract these information. The calculation process of temporal attention is as follows. Firstly, we select a one-channel feature map $F(x,y)$ and apply softmax formula (\ref{attention1}) to calculate the attention maps
\begin{align}\label{attention1}
P_n(x,y)_j=\frac{e^{F_n(x,y)_j}}{\sum_{i=1}^Ne^{F_i(x,y)_j}}, n\in [1:N]
\end{align}
where $P_n(x,y)_j$ represents the weight of the temporal attention task at location $(x,y)_j$. Then, we utilize the following formula to calculate attention-weight feature maps.
\begin{align}\label{attention2}
\overline{F}_n^g=P_n(x,y)_j\otimes F_n^g, n\in [1:N]
\end{align}
where $\overline{F}_n^g$ is the attention-weight feature maps, $F_n^g$ denotes the module to perform the feature attention operation and $\otimes$ indicates element-wise multiplication.
\subsection{Loss function}
In this paper, we utilize L1 loss function for DDCN model. The loss function consists of two parts. The first part considers the loss between the image after bicubic upsampling and the real image. The second part is the loss between the reconstructed image and the real image. The specific loss function has the following form
\begin{align}\label{loss-function}
\nonumber &\mathcal{L}_{UP}=||I_{UP}^H-I_R^H||_1,\\
\nonumber &\mathcal{L}_{IR}=||I_{IR}^H-I_R^H||_1,\\
&\mathcal{L}=\mathcal{L}_{IR}+0.01\times \mathcal{L}_{UP}.
\end{align}
where $\mathcal{L}_{UP}, \mathcal{L}_{IR}$ and $\mathcal{L}$ represent bicubic upsampling loss, image reconstruction loss and total loss respectively. Meanwhile, $I_{UP}^H, I_{IR}^H$ and $I_R^H$ indicate the upsampled image, the reconstructed image and the real image, respectively. This loss calculation method considers the influence of upsampled images on the whole model and can further enhance the convergence ability of the model. We apply this loss function to replace the traditional loss function of $\mathcal{L}_{IR}$, and obtain better effects.
\begin{figure*}[htbp]
\centering
\subfigure{
\begin{minipage}[t]{0.215\linewidth}
\centerline{\includegraphics[width=1.0\textwidth,height=0.75\textwidth]{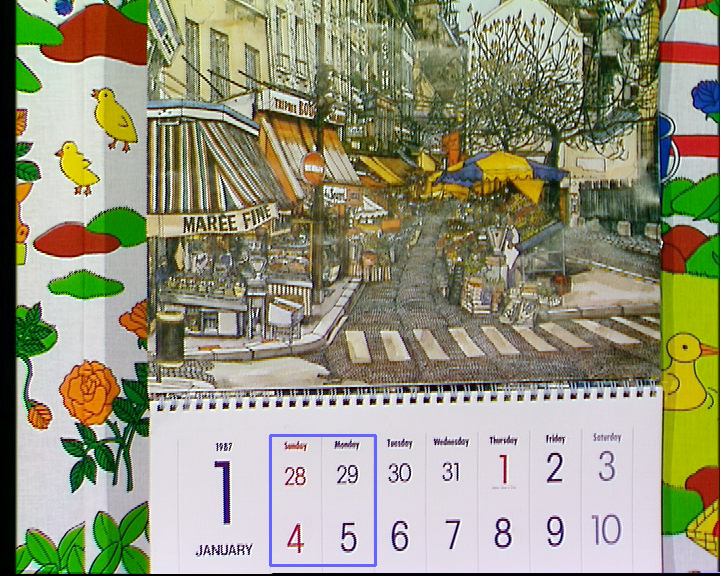}}
\centerline{truth}
\end{minipage}%
}%
\subfigure{
\begin{minipage}[t]{0.215\linewidth}
\centerline{\includegraphics[width=1.0\textwidth,height=0.75\textwidth]{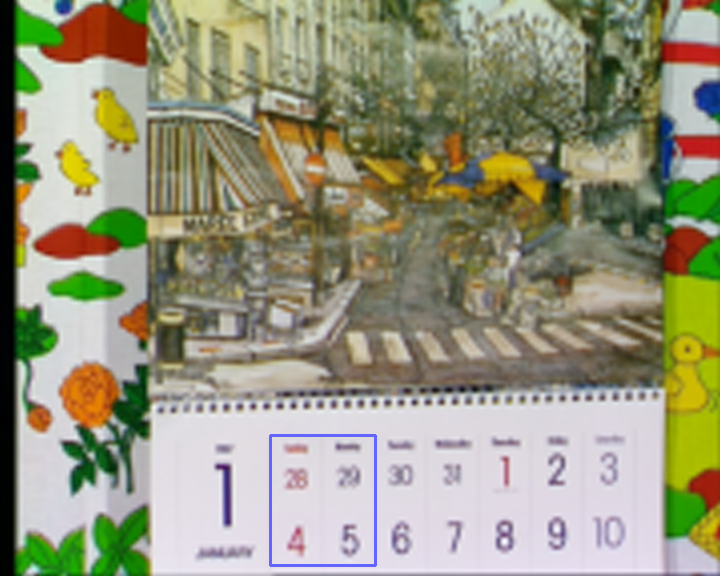}}
\centerline{bicubic}
\end{minipage}%
}%
\subfigure{
\begin{minipage}[t]{0.215\linewidth}
\centerline{\includegraphics[width=1.0\textwidth,height=0.75\textwidth]{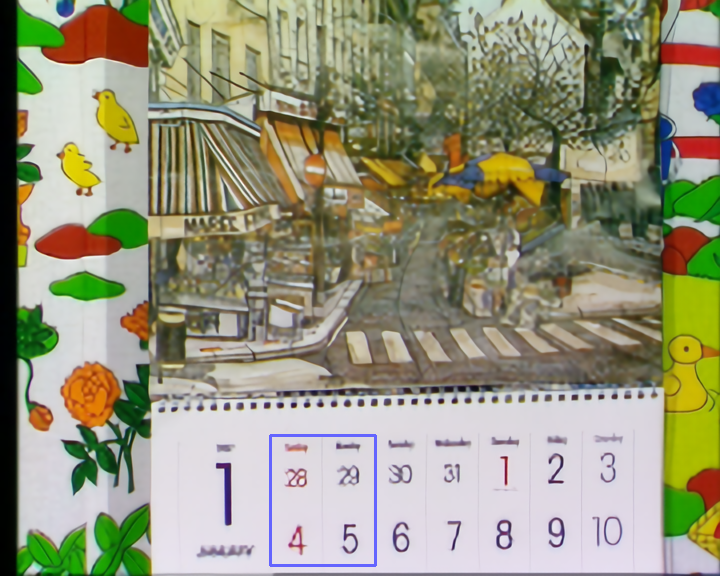}}
\centerline{DBPN}
\end{minipage}%
}%
\subfigure{
\begin{minipage}[t]{0.215\linewidth}
\centerline{\includegraphics[width=1.0\textwidth,height=0.75\textwidth]{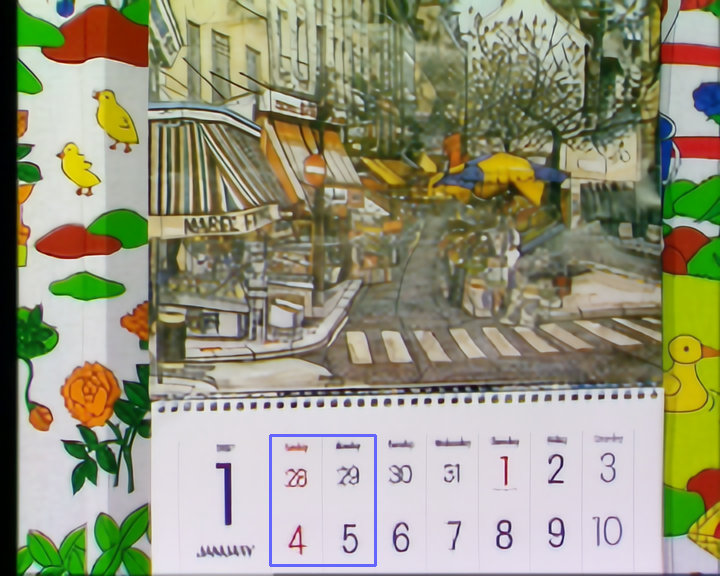}}
\centerline{RCAN}
\end{minipage}%
}%

\subfigure{
\begin{minipage}[t]{0.21\linewidth}
\centerline{\includegraphics[width=1.0\textwidth,height=0.75\textwidth]{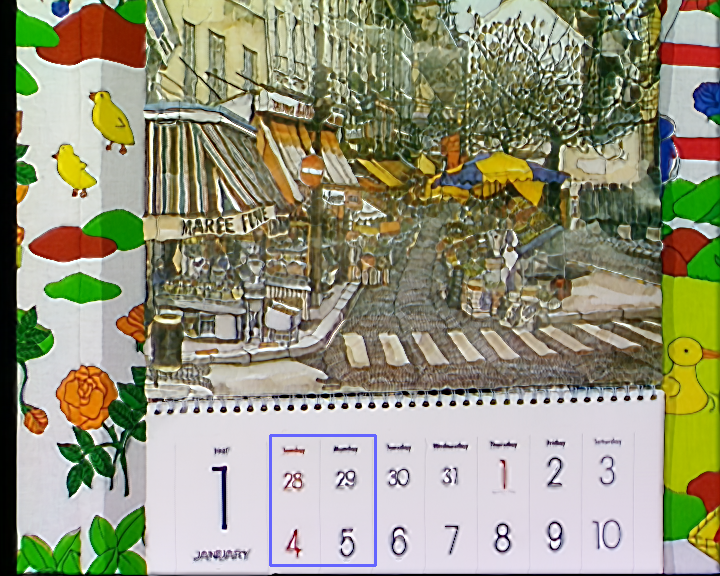}}
\centerline{TOFLOW}
\end{minipage}
}%
\subfigure{
\begin{minipage}[t]{0.21\linewidth}
\centerline{\includegraphics[width=1.0\textwidth,height=0.75\textwidth]{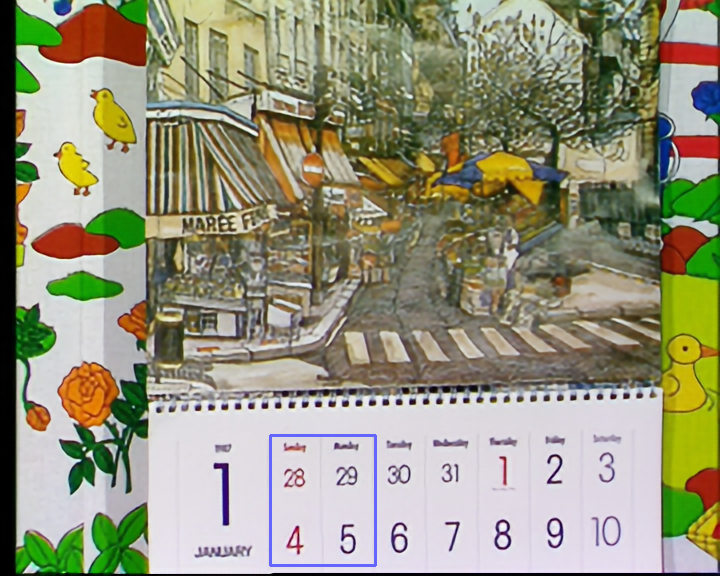}}
\centerline{RISTN}
\end{minipage}
}%
\subfigure{
\begin{minipage}[t]{0.21\linewidth}
\centerline{\includegraphics[width=1.0\textwidth,height=0.75\textwidth]{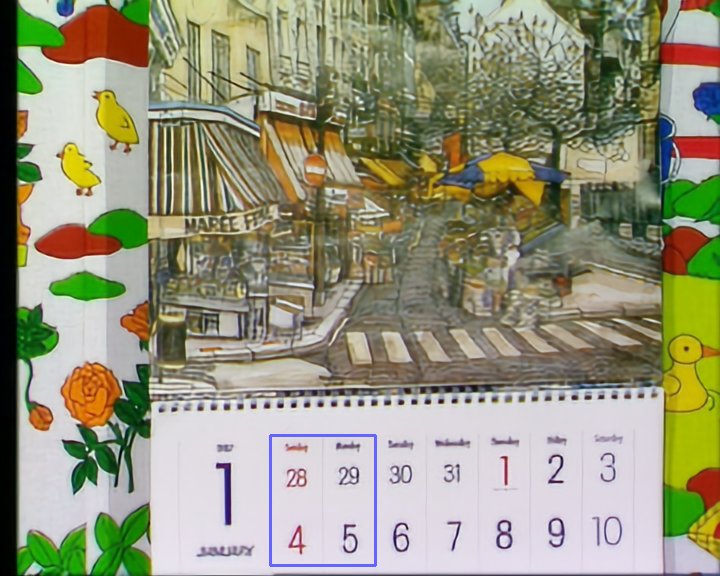}}
\centerline{TDAN}
\end{minipage}
}%
\subfigure{
\begin{minipage}[t]{0.21\linewidth}
\centerline{\includegraphics[width=1.0\textwidth,height=0.75\textwidth]{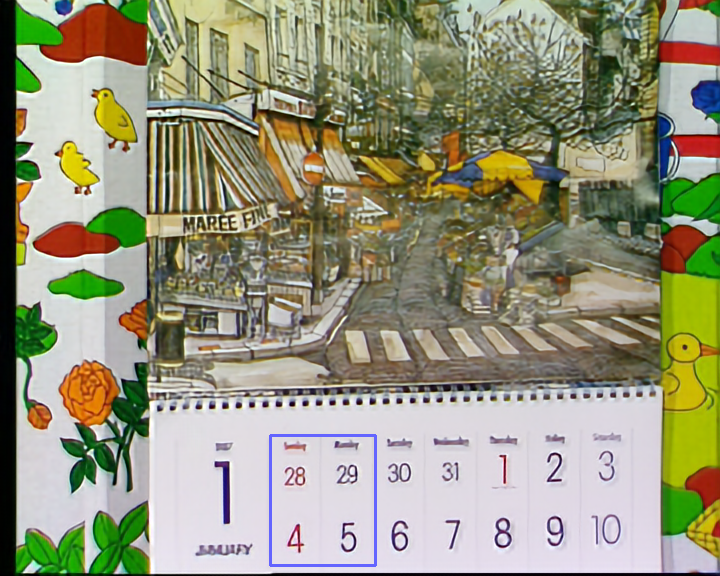}}
\centerline{our}
\end{minipage}
}%

\subfigure{
\begin{minipage}[t]{0.1\linewidth}
\centerline{\includegraphics[width=1.0\textwidth,height=1.1\textwidth]{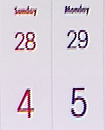}}
\centerline{truth}
\end{minipage}
}%
\subfigure{
\begin{minipage}[t]{0.1\linewidth}
\centerline{\includegraphics[width=1.0\textwidth,height=1.1\textwidth]{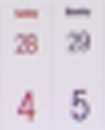}}
\centerline{bicubic}
\end{minipage}
}%
\subfigure{
\begin{minipage}[t]{0.1\linewidth}
\centerline{\includegraphics[width=1.0\textwidth,height=1.1\textwidth]{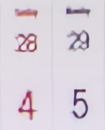}}
\centerline{DBPN}
\end{minipage}
}%
\subfigure{
\begin{minipage}[t]{0.1\linewidth}
\centerline{\includegraphics[width=1.0\textwidth,height=1.1\textwidth]{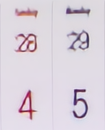}}
\centerline{RCAN}
\end{minipage}
}%
\subfigure{
\begin{minipage}[t]{0.1\linewidth}
\centerline{\includegraphics[width=1.0\textwidth,height=1.1\textwidth]{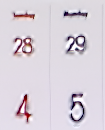}}
\centerline{TOFLOW}
\end{minipage}
}%
\subfigure{
\begin{minipage}[t]{0.1\linewidth}
\centerline{\includegraphics[width=1.0\textwidth,height=1.1\textwidth]{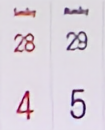}}
\centerline{RISTN}
\end{minipage}
}%
\subfigure{
\begin{minipage}[t]{0.1\linewidth}
\centerline{\includegraphics[width=1.0\textwidth,height=1.1\textwidth]{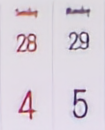}}
\centerline{TDAN}
\end{minipage}
}%
\subfigure{
\begin{minipage}[t]{0.1\linewidth}
\centerline{\includegraphics[width=1.0\textwidth,height=1.1\textwidth]{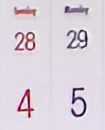}}
\centerline{our}
\end{minipage}
}%
\centering
\caption{Qualitative comparison on the Vid4 for 4$\times$ SR.}
\label{Vid4_image}
\end{figure*}
\section{Experiments}
\subsection{Training Datasets and Details}
\textbf{Datasets} We utilize Vimeo-90K \cite{dataset-Vimeo-90K} as the training set. This dataset includes more than 90K video sequences which each sequence consists of 7 consecutive frames and the frame has a resolution of $448\times256$. Meanwhile, we apply Vid4 \cite{dataset-Vid4} and SPMCS-11 \cite{dataset-SPMCS} as the test sets. Vid4 is a classic dataset which includes four parts: calendar, city, foliage and walk. SPMCS-11 is proposed and accepted in recent years. It contains eleven sequences covering the natural-world and urban scenes, and each sequence consists of 31 consecutive frames with a resolution of $960\times540$.
\begin{figure*}[htbp]
\centering
\subfigure{
\begin{minipage}[t]{0.215\linewidth}
\centerline{\includegraphics[width=1.0\textwidth,height=0.6\textwidth]{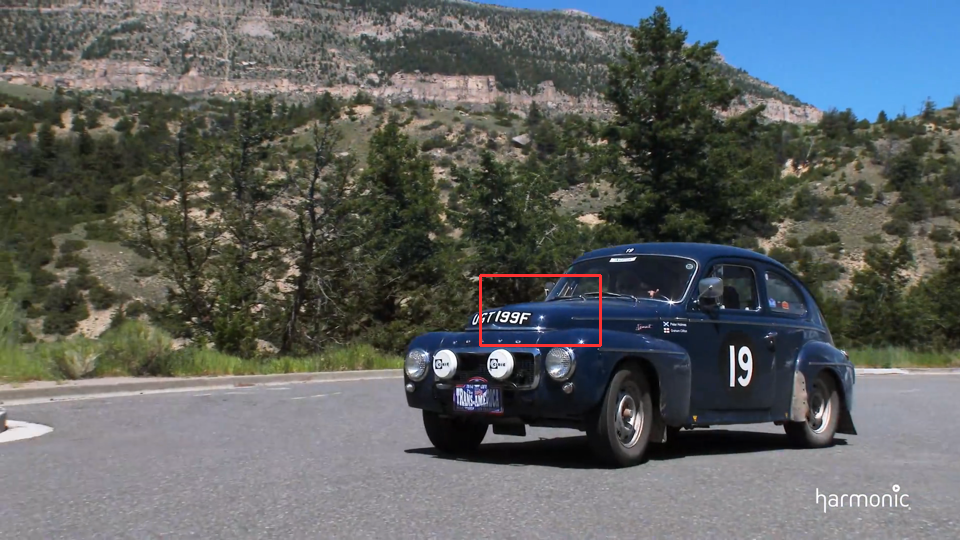}}
\centerline{truth}
\end{minipage}%
}%
\subfigure{
\begin{minipage}[t]{0.215\linewidth}
\centerline{\includegraphics[width=1.0\textwidth,height=0.6\textwidth]{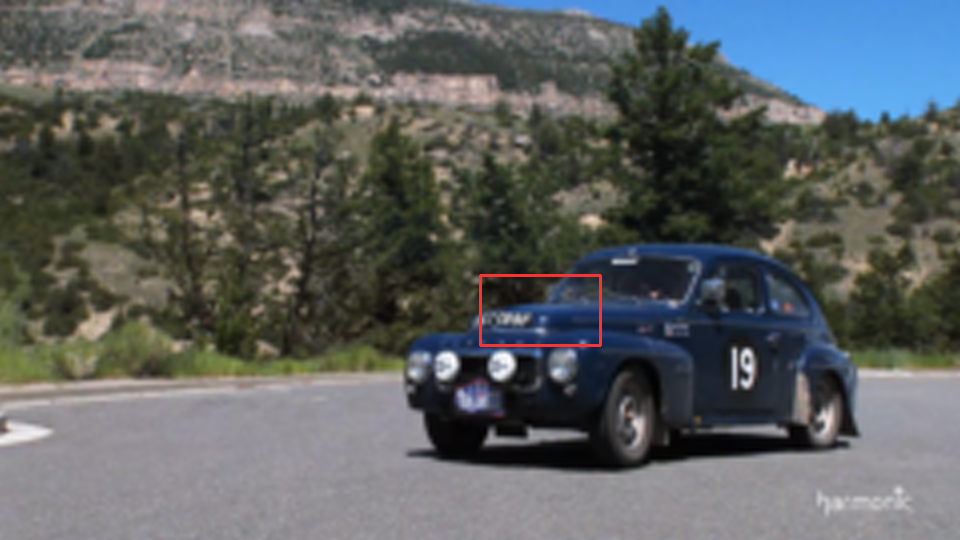}}
\centerline{bicubic}
\end{minipage}%
}%
\subfigure{
\begin{minipage}[t]{0.215\linewidth}
\centerline{\includegraphics[width=1.0\textwidth,height=0.6\textwidth]{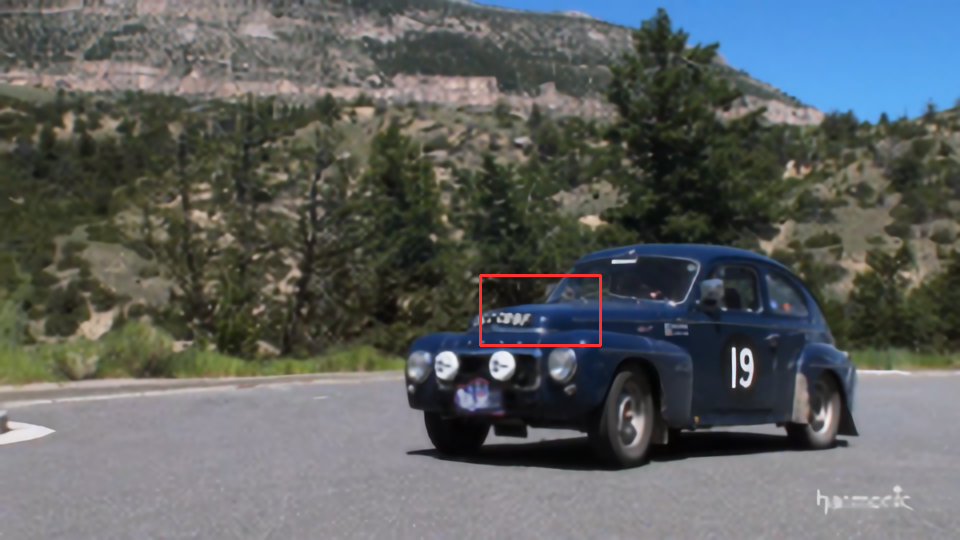}}
\centerline{DBPN}
\end{minipage}%
}%
\subfigure{
\begin{minipage}[t]{0.215\linewidth}
\centerline{\includegraphics[width=1.0\textwidth,height=0.6\textwidth]{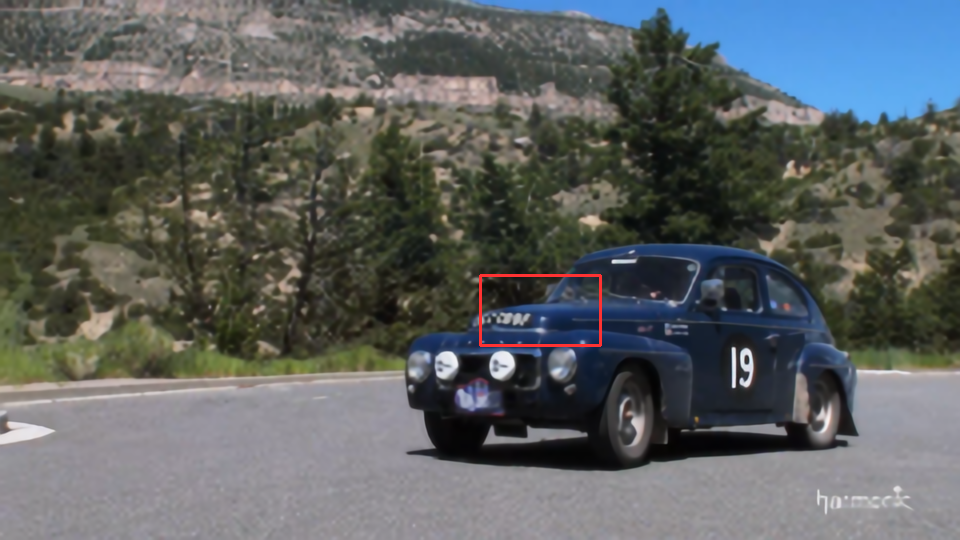}}
\centerline{RCAN}
\end{minipage}%
}%

\subfigure{
\begin{minipage}[t]{0.21\linewidth}
\centerline{\includegraphics[width=1.0\textwidth,height=0.6\textwidth]{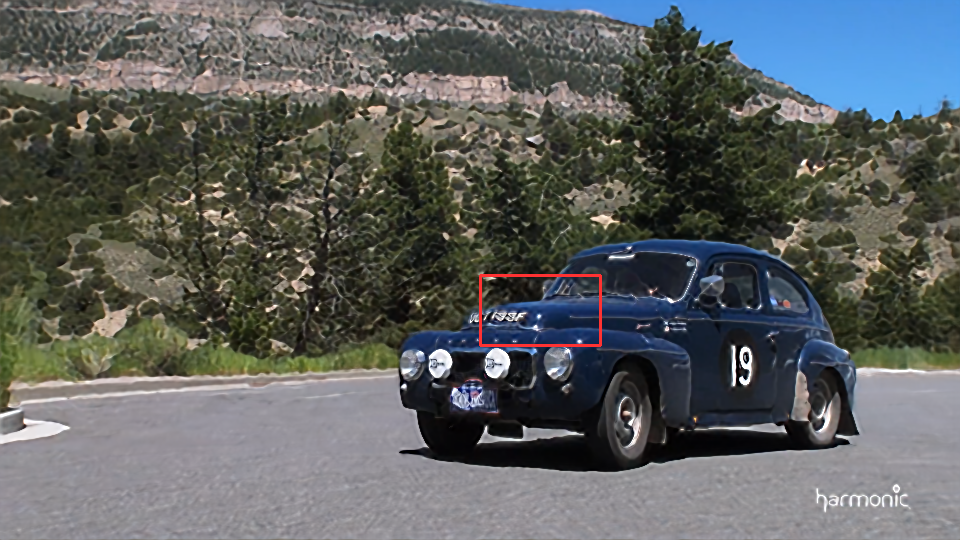}}
\centerline{TOFLOW}
\end{minipage}
}%
\subfigure{
\begin{minipage}[t]{0.21\linewidth}
\centerline{\includegraphics[width=1.0\textwidth,height=0.6\textwidth]{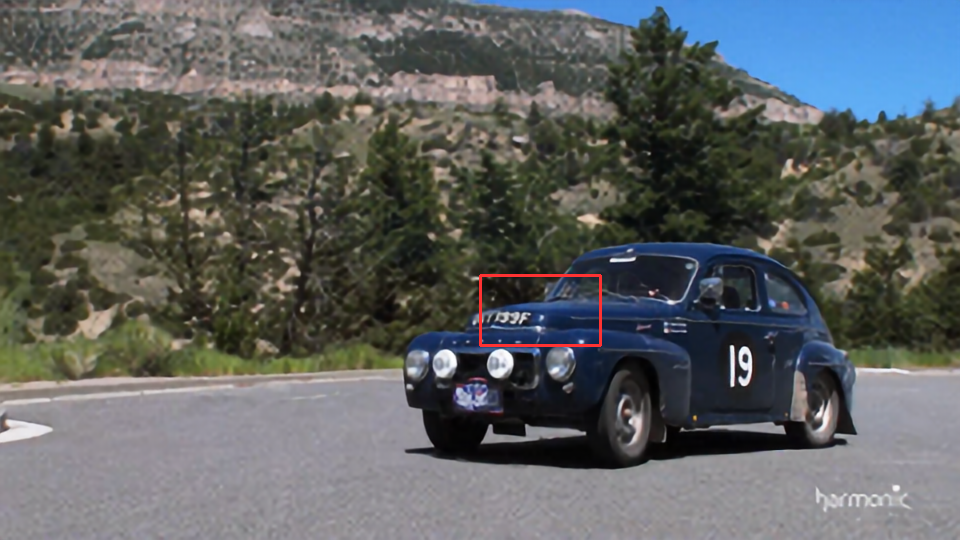}}
\centerline{RISTN}
\end{minipage}
}%
\subfigure{
\begin{minipage}[t]{0.21\linewidth}
\centerline{\includegraphics[width=1.0\textwidth,height=0.6\textwidth]{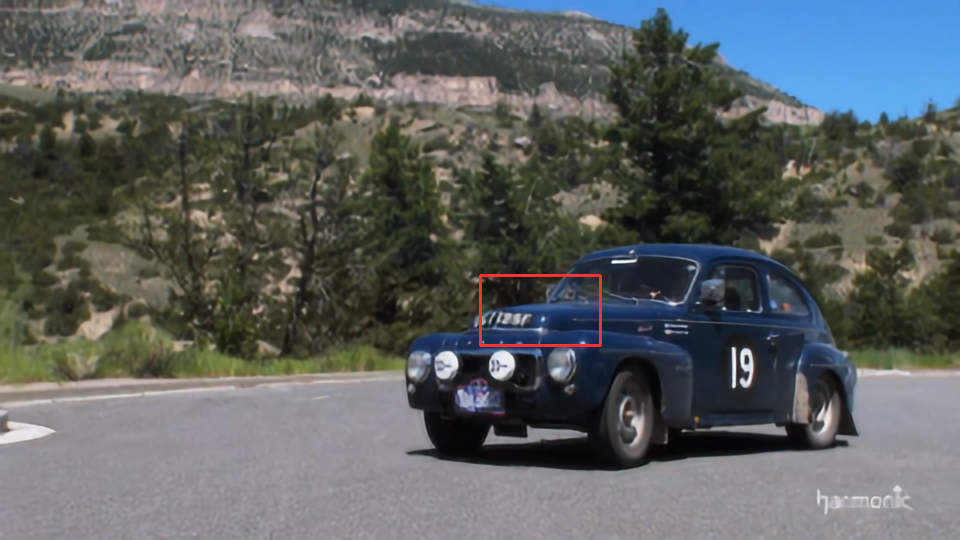}}
\centerline{TDAN}
\end{minipage}
}%
\subfigure{
\begin{minipage}[t]{0.21\linewidth}
\centerline{\includegraphics[width=1.0\textwidth,height=0.6\textwidth]{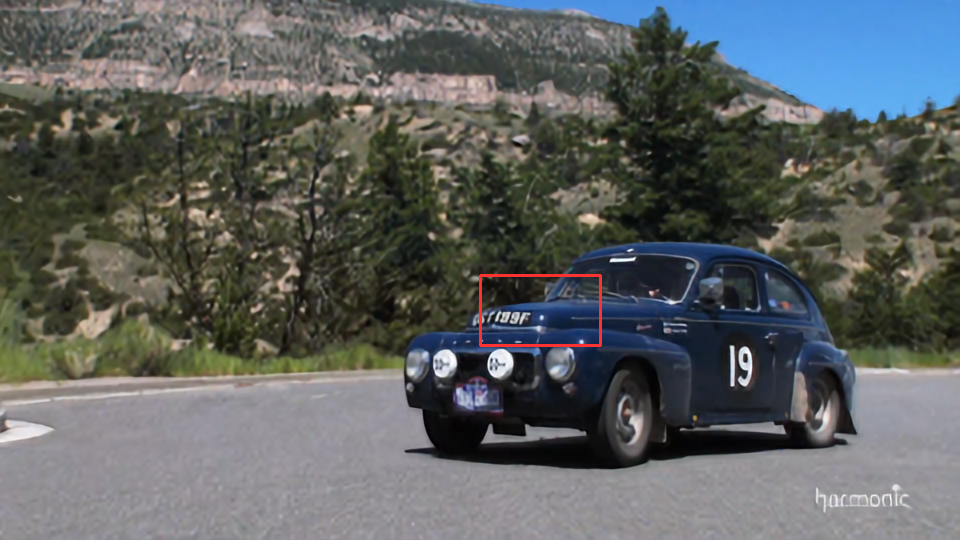}}
\centerline{our}
\end{minipage}
}%

\subfigure{
\begin{minipage}[t]{0.1\linewidth}
\centerline{\includegraphics[width=1.0\textwidth,height=0.6\textwidth]{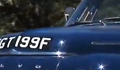}}
\centerline{truth}
\end{minipage}
}%
\subfigure{
\begin{minipage}[t]{0.1\linewidth}
\centerline{\includegraphics[width=1.0\textwidth,height=0.6\textwidth]{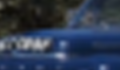}}
\centerline{bicubic}
\end{minipage}
}%
\subfigure{
\begin{minipage}[t]{0.1\linewidth}
\centerline{\includegraphics[width=1.0\textwidth,height=0.6\textwidth]{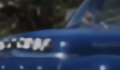}}
\centerline{DBPN}
\end{minipage}
}%
\subfigure{
\begin{minipage}[t]{0.1\linewidth}
\centerline{\includegraphics[width=1.0\textwidth,height=0.6\textwidth]{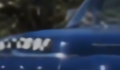}}
\centerline{RCAN}
\end{minipage}
}%
\subfigure{
\begin{minipage}[t]{0.1\linewidth}
\centerline{\includegraphics[width=1.0\textwidth,height=0.6\textwidth]{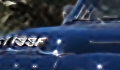}}
\centerline{TOFLOW}
\end{minipage}
}%
\subfigure{
\begin{minipage}[t]{0.1\linewidth}
\centerline{\includegraphics[width=1.0\textwidth,height=0.6\textwidth]{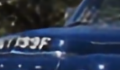}}
\centerline{RISTN}
\end{minipage}
}%
\subfigure{
\begin{minipage}[t]{0.1\linewidth}
\centerline{\includegraphics[width=1.0\textwidth,height=0.6\textwidth]{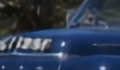}}
\centerline{TDAN}
\end{minipage}
}%
\subfigure{
\begin{minipage}[t]{0.1\linewidth}
\centerline{\includegraphics[width=1.0\textwidth,height=0.6\textwidth]{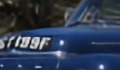}}
\centerline{our}
\end{minipage}
}%
\centering
\caption{Qualitative comparison on the SPMCS-11 for 4$\times$ SR.}
\label{SPMCS_image1}
\end{figure*}

\textbf{Implementation details} For training set, we crop the HR image to $256\times256$. The size of LR image is $64\times64$ by applying Gaussian blur with the standard deviation of $\sigma=1.6$ and $4\times$ downsampling. In model training, we utilize Adam optimizer with $\beta_1=0.9$ and $\beta_2=0.999$. The learning rate is initially set to $1\times10^{-4}$ and dropped to $1\times10^{-5}$ after 40 batches. Then loop 15 times with a learning rate of $1\times10^{-5}$. The batch size is 8. Moreover, in order to increase the range of the training set, we randomly flip the input frames horizontally and vertically.
\begin{figure*}[htbp]
\centering
\subfigure{
\begin{minipage}[t]{0.215\linewidth}
\centerline{\includegraphics[width=1.0\textwidth,height=0.6\textwidth]{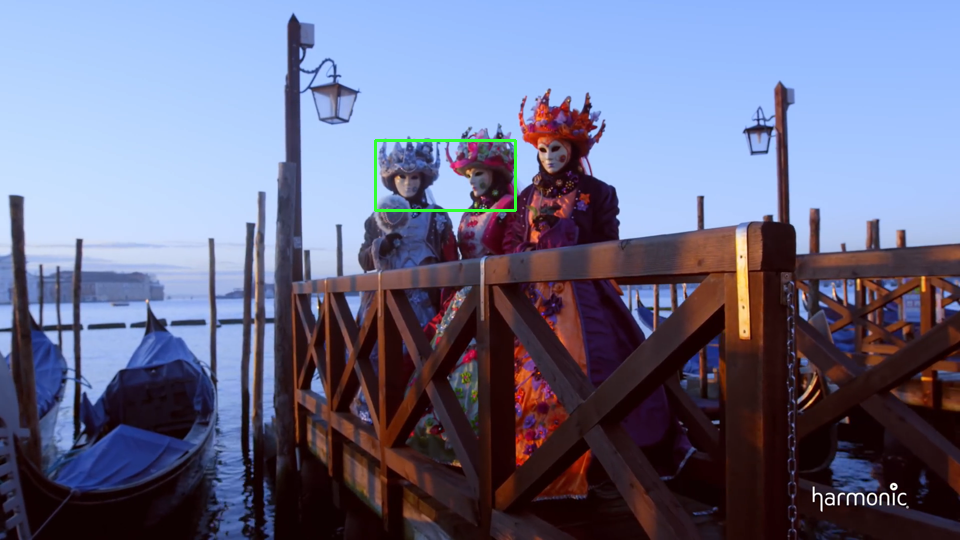}}
\centerline{truth}
\end{minipage}%
}%
\subfigure{
\begin{minipage}[t]{0.215\linewidth}
\centerline{\includegraphics[width=1.0\textwidth,height=0.6\textwidth]{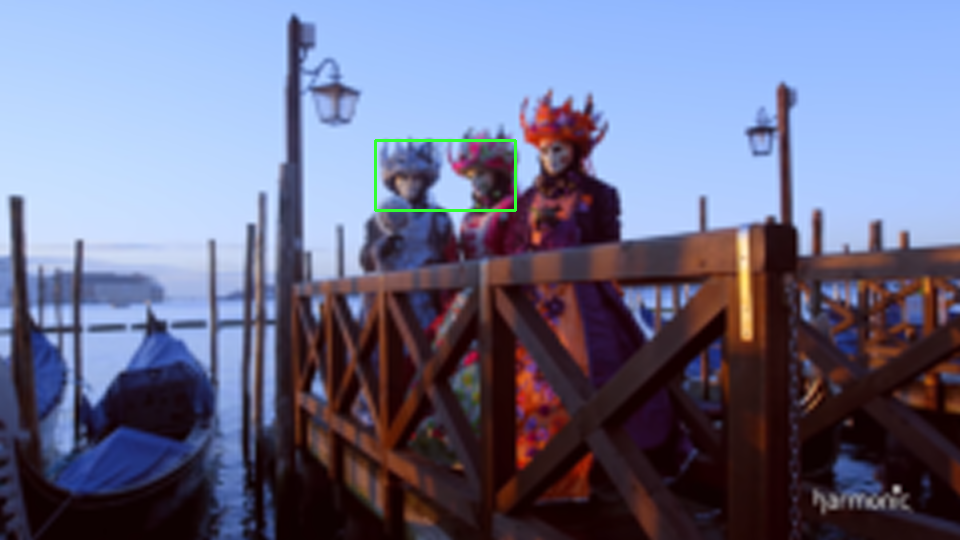}}
\centerline{bicubic}
\end{minipage}%
}%
\subfigure{
\begin{minipage}[t]{0.215\linewidth}
\centerline{\includegraphics[width=1.0\textwidth,height=0.6\textwidth]{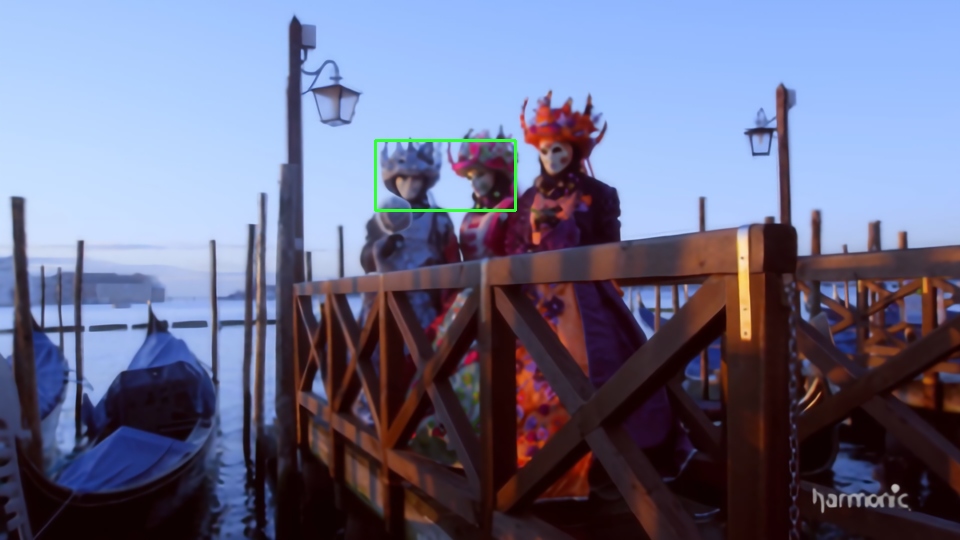}}
\centerline{DBPN}
\end{minipage}%
}%
\subfigure{
\begin{minipage}[t]{0.215\linewidth}
\centerline{\includegraphics[width=1.0\textwidth,height=0.6\textwidth]{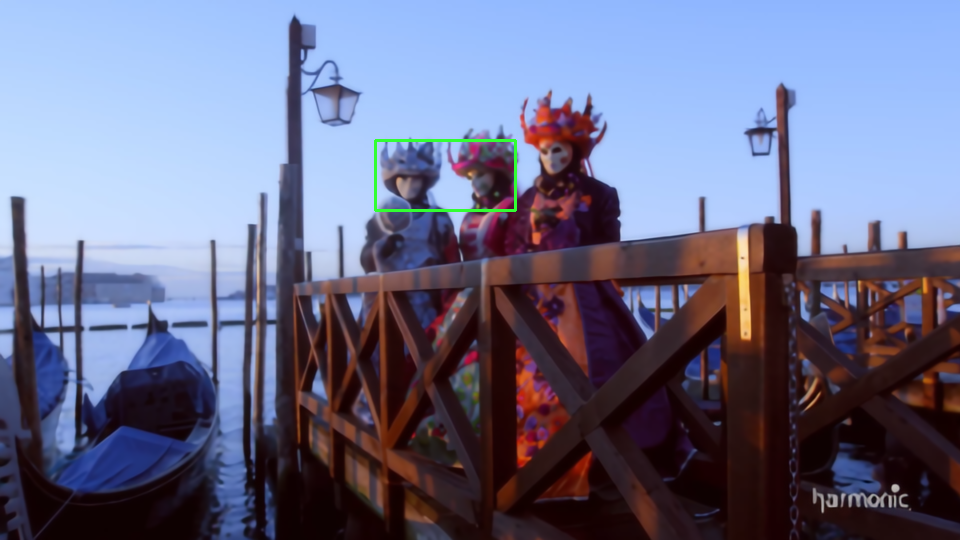}}
\centerline{RCAN}
\end{minipage}%
}%

\subfigure{
\begin{minipage}[t]{0.21\linewidth}
\centerline{\includegraphics[width=1.0\textwidth,height=0.6\textwidth]{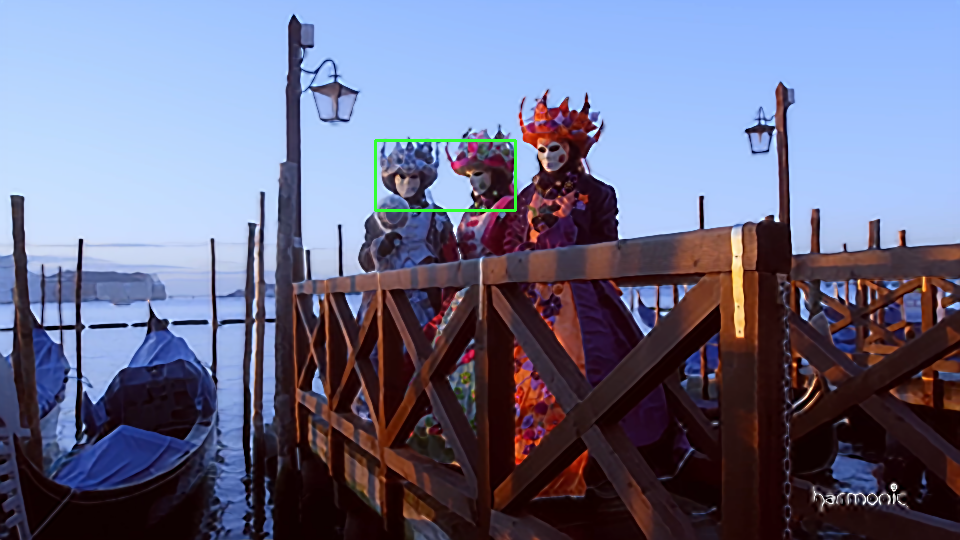}}
\centerline{TOFLOW}
\end{minipage}
}%
\subfigure{
\begin{minipage}[t]{0.21\linewidth}
\centerline{\includegraphics[width=1.0\textwidth,height=0.6\textwidth]{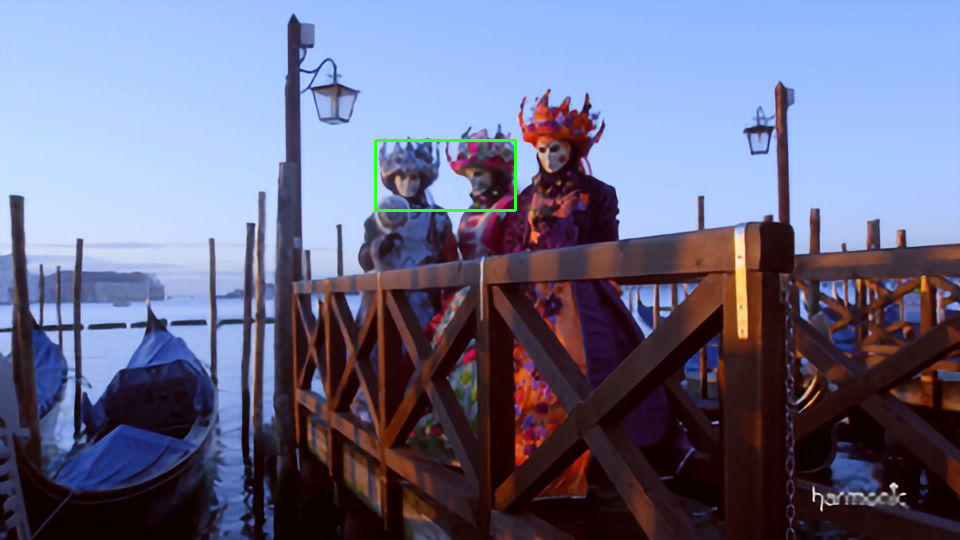}}
\centerline{RISTN}
\end{minipage}
}%
\subfigure{
\begin{minipage}[t]{0.21\linewidth}
\centerline{\includegraphics[width=1.0\textwidth,height=0.6\textwidth]{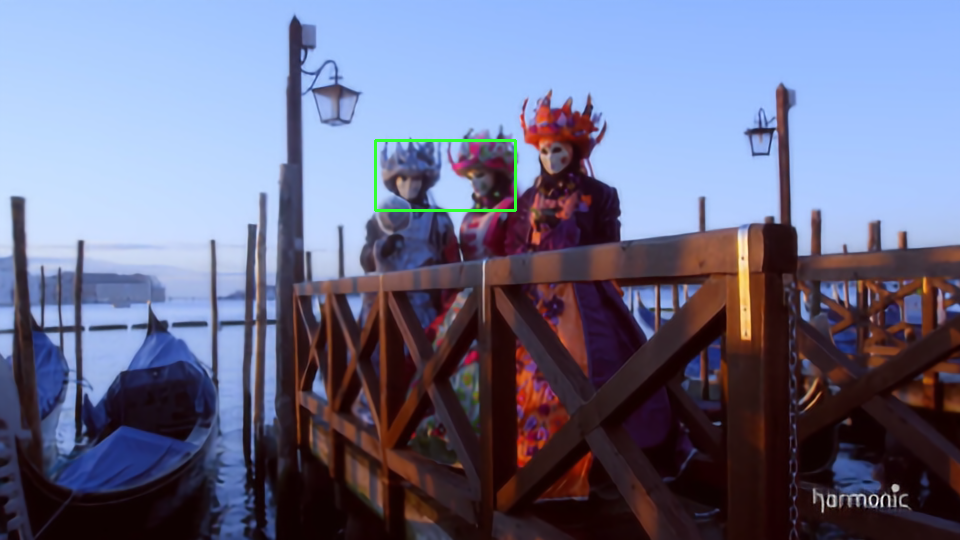}}
\centerline{TDAN}
\end{minipage}
}%
\subfigure{
\begin{minipage}[t]{0.21\linewidth}
\centerline{\includegraphics[width=1.0\textwidth,height=0.6\textwidth]{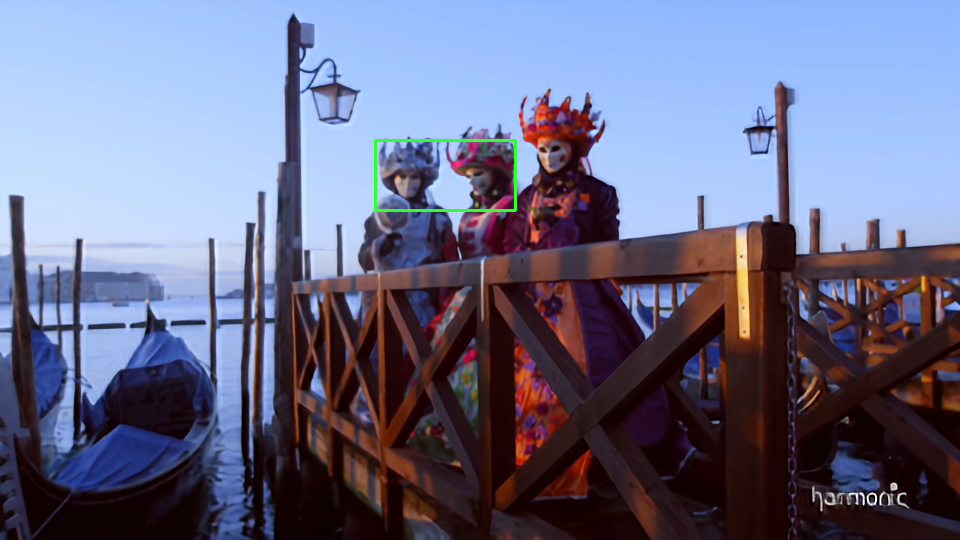}}
\centerline{our}
\end{minipage}
}%

\subfigure{
\begin{minipage}[t]{0.1\linewidth}
\centerline{\includegraphics[width=1.0\textwidth,height=0.65\textwidth]{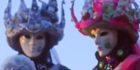}}
\centerline{truth}
\end{minipage}
}%
\subfigure{
\begin{minipage}[t]{0.1\linewidth}
\centerline{\includegraphics[width=1.0\textwidth,height=0.65\textwidth]{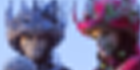}}
\centerline{bicubic}
\end{minipage}
}%
\subfigure{
\begin{minipage}[t]{0.1\linewidth}
\centerline{\includegraphics[width=1.0\textwidth,height=0.65\textwidth]{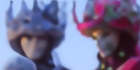}}
\centerline{DBPN}
\end{minipage}
}%
\subfigure{
\begin{minipage}[t]{0.1\linewidth}
\centerline{\includegraphics[width=1.0\textwidth,height=0.65\textwidth]{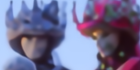}}
\centerline{RCAN}
\end{minipage}
}%
\subfigure{
\begin{minipage}[t]{0.1\linewidth}
\centerline{\includegraphics[width=1.0\textwidth,height=0.65\textwidth]{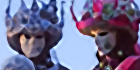}}
\centerline{TOFLOW}
\end{minipage}
}%
\subfigure{
\begin{minipage}[t]{0.1\linewidth}
\centerline{\includegraphics[width=1.0\textwidth,height=0.65\textwidth]{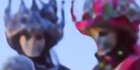}}
\centerline{RISTN}
\end{minipage}
}%
\subfigure{
\begin{minipage}[t]{0.1\linewidth}
\centerline{\includegraphics[width=1.0\textwidth,height=0.65\textwidth]{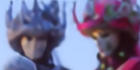}}
\centerline{TDAN}
\end{minipage}
}%
\subfigure{
\begin{minipage}[t]{0.1\linewidth}
\centerline{\includegraphics[width=1.0\textwidth,height=0.65\textwidth]{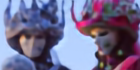}}
\centerline{our}
\end{minipage}
}%
\centering
\caption{Qualitative comparison on the SPMCS-11 for 4$\times$ SR.}
\label{SPMCS_image2}
\end{figure*}

\begin{table*}[htp]
\centering
%\caption{}\label{results}
\begin{tabular}{c|c|c|c|c|c|c|c}
\multicolumn{8}{l}{\small{\textbf{Table 1}}}\\
\multicolumn{8}{l}{\small{Quantitative comparison (PSNR(dB) and SSIM) on Vid4 and SPMCS-11 for 4$\times$. The PSNR and SSIM are calculated on the Y-channel.}}\\
\multicolumn{8}{l}{\small{Red texts indicate the best result.}}\\
\hline
Method & Bicubic & DBPN & RCAN & TOFLOW & RISTN & TDAN & Ours\\
\hline
Vid4 & 21.80/0.542 & 25.36/0.737 & 25.43/0.738 & 25.85/0.767 & 26.15/0.791 & 26.18/0.783 & \textcolor{red}{26.34}/\textcolor{red}{0.799} \\
\hline
SPMCS-11 & 23.29/0.638 & 28.10/0.820 & 28.36/0.828 & 27.86/0.824 & 28.36/0.836 & 28.39/0.841 & \textcolor{red}{28.40}/\textcolor{red}{0.846} \\
\hline
\end{tabular}
\end{table*}

%\begin{table*}[htp]
%\centering
%\footnotesize \caption{Quantitative comparison (PSNR(dB) and SSIM) on Vid4 and SPMCS-11 for 4$\times$. The PSNR and SSIM are calculated on the Y-channel. Red texts indicate the best result.}
%\label{results}
%\begin{tabular}{c|c|c|c|c|c|c|c}
%\hline
%Method & Bicubic & DBPN & RCAN & TOFLOW & RISTN & TDAN & Ours\\
%\hline
%Vid4 & 21.80/0.542 & 25.36/0.737 & 25.43/0.738 & 25.85/0.767 & 26.15/0.791 & 26.18/0.783 & \textcolor{red}{26.34}/\textcolor{red}{0.799} \\
%\hline
%SPMCS-11 & 23.29/0.638 & 28.10/0.820 & 28.36/0.828 & 27.86/0.824 & 28.36/0.836 & 28.39/0.841 & \textcolor{red}{28.40}/\textcolor{red}{0.846} \\
%\hline
%\end{tabular}
%\end{table*}
\subsection{Comparison with State-of-the-art Methods}
We compare our model with current advanced models including DBPN \cite{Eval-DBPN}, RCAN \cite{Eval-RCAN}, TOFLOW \cite{dataset-Vimeo-90K}, RISTN \cite{Eval-RISTN} and TDAN \cite{deformable2D}. In order to ensure the fairness of comparison, we carefully implement DBPN, TOFLOW, RISTN and TDAN, and rebuild RCAN with the public code. We have trained these models on the same training datasets and the same downsampling method. We record the quantitative comparisons in Table 1. These results come from the original paper or calculated by ourselves. Compared with these algorithms, our model possesses obvious advantages.

\textbf{Objective results} On the Vid4 datasets, depending on Table 1, we can clearly know that our PSNR is much higher than DBPN, RCAN and TOFLOW. Meanwhile, our PSNR outperforms RISTN and TDAN by 0.19dB and 0.16dB respectively. On SSIM, our model obtain better results than these models. On the SPMCS-11 datasets, our model also outperforms other models on both PSNR and SSIM.

\textbf{Visualization} In order to acquire an intuitive feeling effect, we have displayed the test effects of different models. In \textbf{Fig. \ref{Vid4_image}}, we select an image from the calendar in Vid4 datasets. Through comparison, it can be found that DDCN model has better detail recovery ability, and can clearly restore the digit on the calendar. In \textbf{Fig. \ref{SPMCS_image1}} and \textbf{Fig. \ref{SPMCS_image2}}, we choose two images of SPMCS-11 datasets. Compared with image produced by other models, the image reconstructed by DDCN model possesses sharper edges and finer details.
\subsection{Ablation Studies}
In this section, we conduct ablation studies on temporal frames, temporal attention and loss function respectively. The purpose is to further prove the validity of the model.

\textbf{Temporal groups} In this paper, we put forward pre-temporal group and post-temporal group, which represent we can enter any odd number of frames. It is easy to notice that different numbers of input frames can produce different results. To prove that DDCN has the most advanced performance when the number of input frames is 5, we set the number of input frames to 3, 5, and 7 and compare their result. The comparison results are shown in Table 2.

\begin{table}[htp]
\centering
%\caption{}\label{input_frames}
\begin{tabular}{c|c|c|c}
\multicolumn{4}{l}{\small{\textbf{Table 2}}}\\
\multicolumn{4}{l}{\small{The result of different input frames.}}\\
\hline
Frames & 3 & 5 & 7\\
\hline
Vid4 & 25.98/0.770 & 26.34/0.799 & 26.11/0.783\\
\hline
SPMCS-11 & 27.93/0.812 & 28.40/0.846 & 28.06/0.831\\
\hline
\end{tabular}
\end{table}

%\begin{table}[htp]
%\centering
%\footnotesize \caption{The result of different input frames.}
%\label{input_frames}
%\begin{tabular}{c|c|c|c}
%\hline
%Frames & 3 & 5 & 7\\
%\hline
%Vid4 & 25.98/0.770 & 26.34/0.799 & 26.11/0.783\\
%\hline
%SPMCS-11 & 27.93/0.812 & 28.40/0.846 & 28.06/0.831\\
%\hline
%\end{tabular}
%\end{table}

In Table 2, we find when the number of input frames is 5, the acquired result is the highest. PSNR is 0.36dB and 0.23dB higher than input frames 3 and 7 on Vid4 datasets, respectively. This indicates that when the number of input frames is small, the reference frame occupies too much weight, and sufficient temporal information can not be extracted. When the number of input frames is large, the information of other frames is too much, which disturbs the super resolution reconstruction of the current frame. In DDCN, the best evaluation effect can be acquired when the number of input frames is 5.

\textbf{Temporal Attention}
In feature fusion module and feature reconstruction module, we apply the temporal attention mechanism to further extract the features. To demonstrate the effect of temporal attention, we remove the attention mechanism of feature extraction module and feature fusion module respectively. The evaluation results are displayed in Table 3.

\begin{table}[htp]
\centering
%\caption{} \label{att_results}
\begin{tabular}{c|c|c|c}
\multicolumn{4}{l}{\small{\textbf{Table 3}}}\\
\multicolumn{4}{l}{\small{Temporal Attention Results. In this table, FEA and FFM represent}}\\
\multicolumn{4}{l}{\small{feature extraction module and feature fusion module without}}\\
\multicolumn{4}{l}{\small{attention mechanism respectively.}}\\
\hline
 & FEA & FFM & None\\
\hline
Vid4 & 25.61/0.774 & 25.83/0.783 & 26.34/0.799\\
\hline
SPMCS-11 & 27.98/0.813 & 28.08/0.832 & 28.40/0.846\\
\hline
\end{tabular}
\end{table}

%\begin{table}[htp]
%\centering
%\footnotesize \caption{Temporal Attention Results. In this table, FEA and FFM represent feature extraction module and feature fusion module respectively.}
%\label{att_results}
%\begin{tabular}{c|c|c|c}
%\hline
% & FEA & FFM & None\\
%\hline
%Vid4 & 25.61/0.774 & 25.83/0.783 & 26.34/0.799\\
%\hline
%SPMCS-11 & 27.98/0.813 & 28.08/0.832 & 28.40/0.846\\
%\hline
%\end{tabular}
%\end{table}

In Table 3, we notice that whether the attention mechanism of feature extraction module is removed or the attention mechanism of the feature fusion module is removed, the obtained effect  is lower than the model with the attention mechanism added. This indicates that the attention mechanism plays a positive role in DDCN model.

\textbf{Loss function}
In this paper, we design a novel loss function. To prove the validity of this loss function, we design DDCN model using traditional loss function. The comparison results are shown in Table 4.

\begin{table}[htp]
\centering
%\caption{} \label{loss_function_results}
\begin{tabular}{c|c|c}
\multicolumn{3}{l}{\small{\textbf{Table 4}}}\\
\multicolumn{3}{l}{\small{Model results under different loss functions. $\mathcal{L}_{IR}$ indicates the }}\\
\multicolumn{3}{l}{\small{traditional loss function. $\mathcal{L}$ represents the new loss function.}}\\
\multicolumn{3}{l}{\small{The symbols of $\mathcal{L}_{IR}$ and $\mathcal{L}$ are explained in (\ref{loss-function}).}}\\

\hline
 & $\mathcal{L}_{IR}$ & $\mathcal{L} $\\
\hline
Vid4 & 26.25/0.781 & 26.34/0.799\\
\hline
SPMCS-11 & 28.22/0.831 & 28.40/0.846\\
\hline
\end{tabular}
\end{table}

%\begin{table}[htp]
%\centering
%\footnotesize \caption{Model results under different loss functions. $\mathcal{L}_{IR}$ indicates the traditional loss function. $\mathcal{L}$ represents the new loss function. The symbols of $\mathcal{L}_{IR}$ and $\mathcal{L}$ are explained in (\ref{loss-function}).}
%\label{loss_function_results}
%\begin{tabular}{c|c|c}
%\hline
% & $\mathcal{L}_{IR}$ & $\mathcal{L} $\\
%\hline
%Vid4 & 26.25/0.781 & 26.34/0.799\\
%\hline
%SPMCS-11 & 28.22/0.831 & 28.40/0.846\\
%\hline
%\end{tabular}
%\end{table}

In Table 4, we can clearly understand that the DDCN model achieves better results by applying the new loss function. This indicates that the proposed loss function is meaningful. Adding the loss between the image after bicubic upsampling and the real image to the total loss is helpful to enhance the performance of the model.
\section{Conclusion}
In this paper, we present a novel video super-resolution network which utilize densely connected convolutional network in both inner and outer layers. This method reduces the number of feature maps of the network model while reusing features. Meanwhile, we divide the input sequence into reference frame, pre-temporal group and post-temporal group. This method can prevent temporal information disorder. In addition, we propose a novel loss function method to improve the performance of the model. Experiments on Vid4 and SPMCS-11 demonstrate that our proposed method is effective and possess extensive research value.
\section*{Acknowledgement}
This paper is supported by the National Natural Science Foundation of China(No. 11571325) and the Fundamental Research Funds for the Central Universities(No. CUC2019 A002).

\begin{wrapfigure}{l}{25mm}
\includegraphics[width=1in,height=1.25in,clip,keepaspectratio]{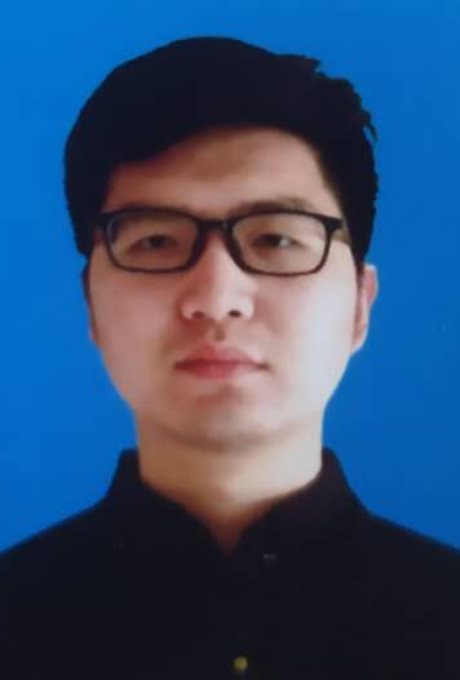}
\end{wrapfigure}\par
\textbf{Guofang Li} graduated from North China University of Technology with a master's degree in mathematics, Beiging, China, in 2020. He is a doctoral student of Communication and Information Systems at Communication University of China. His research interests are machine learning and computer vision.

~\\

\begin{wrapfigure}{l}{25mm}
\includegraphics[width=1in,height=1.25in,clip,keepaspectratio]{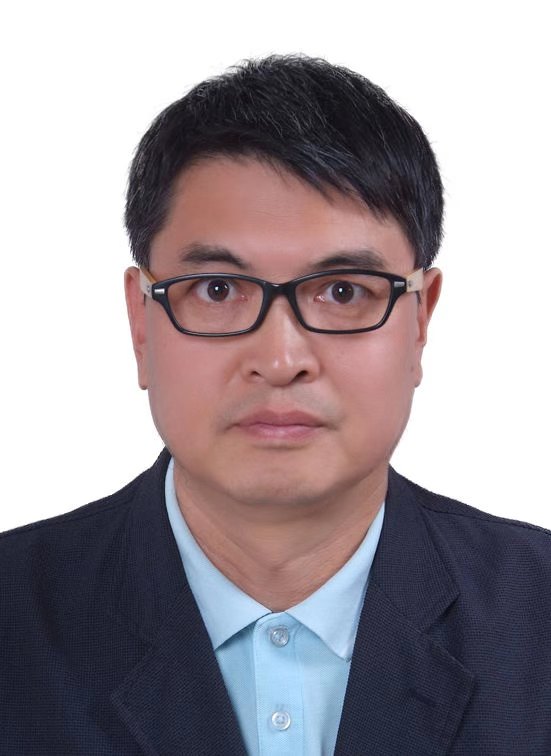}
\end{wrapfigure}\par
\textbf{Yonggui Zhu} received the Ph. D. degree in Academy of Mathematics and Systems Science, Chinese Academy of Sciences, China. He is currently a Professor with the School of Data Science and Media Intelligence, Communication University of China, China. His current research interests include Machine Learning, Deep Learning, Artificial Intelligence, Video Processing, and Software Architecture.

\end{document}